\documentclass{aa}
\usepackage{txfonts}
\usepackage{graphicx}
\begin{document}
 \title{Cold dust and its relation to molecular gas in the dwarf
irregular galaxy \object{NGC~4449}}

 \author{C. B\"ottner\inst{}
  \and U. Klein\inst{}
  \and A. Heithausen\inst{}
   }

 \institute{Radioastronomisches Institut der Universit\"at Bonn, Auf dem
H\"ugel 71, 53121 Bonn, Germany}

 \offprints{\\C.~B\"ottner, \email{cboettne@astro.uni-bonn.de}}

 \date{Received 05 Nov 2002 / Accepted 24 Jun 03}

 \abstract{ We present observations of the dwarf irregular galaxy
 \object{NGC\,4449} at~850 $\mu$m and 450~$\mu$m obtained with SCUBA
 at the JCMT. The distribution of the cold dust agrees well with that
 of the CO and H$\alpha$ emission. To explain the integrated mm- through
 far-infrared continuum spectrum three dust components are required,
 with temperatures of 16~K, 39~K and 168~K, respectively. The dust mass is
dominated by the cold dust component; we derive a total dust mass of
 $\sim3.8~\times~10^6~${M$_\odot$}, and with the local gas-to-dust ratio
 of $\sim130$ a total gas mass of $M_{\rm HI+H_2}\sim4.9~\times~10^8$~{M$_\odot$}.
Comparison with the HI mass leads to a total molecular gas mass of
 $\sim3.4~\times~10^8$~{M$_\odot$}. We derive a conversion factor of the
CO line intensity to molecular hydrogen column density
\mbox{$X_{\rm CO}=N_{\rm H_2}/I_{\rm CO}$} which is at least
11 times larger than the Galactic value.  These values are in accord
with the lower metallicity of \object{NGC\,4449}.

\keywords{galaxies: individual: \object{NGC\,4449} -- galaxies: ISM --
galaxies: irregular -- ISM: dust, extinction}  }

\titlerunning{Cold dust in NGC~4449}
 \maketitle

%
 \section{Introduction}

Carbon monoxide (CO) is commonly used as a tracer for molecular gas
because the molecular hydrogen (H$_2$), the dominant species, lacks
strong emission lines from which the column density could be
determined.  Because the rotational transitions of CO are relatively
easy to excite, it is possible to use the CO line intensity integrated
over all velocities $(I_{\rm CO}=\int T_{\rm mb} dv$) or the CO
luminosity, $L_{\rm CO}$, in one of these lines to estimate the column
density, $N({\rm H_2)}$, and mass, $M({\rm H_2)}$, of molecular
hydrogen, provided one knows the correct conversion. For the Milky Way
the conversion factor, \mbox{$X_{\rm CO}=N_{\rm H_2}/I_{\rm CO(1-0)}$}, has been
found to be between
\mbox{$X_{\rm CO}=1.6\times10^{20}$~cm$^{-2}$~(K~km~s$^{-1})^{-1}$}
(Hunter et al. \cite{1997ApJ...481..205H}) and
\mbox{$X_{\rm CO}=2.3\times10^{20}$~cm$^{-2}$~(K~km~s$^{-1})^{-1}$} (Strong
et al. \cite{strong}).  Throughout the paper we will use the latter
because it is widely used for comparison in the literature.  Early on,
the question arose whether one can apply this local value to external
galaxies, as the relation may strongly depend on the physical
conditions such as the radiation field and the metallicity (Dettmar \&
Heithausen \cite{dettmar}; Israel \cite{israel}, Barone et
al. \cite{barone}). Dwarf irregular galaxies provide excellent
laboratories to investigate {$X_{\rm CO}$} in environments with lower metal
abundances and varying excitation conditions.

An alternative tracer for the molecular gas is the cold dust that one
can measure at mm/submm wavelengths
(Gu\'elin et al. \cite{guelin}; Dumke et al. \cite{dumke}).
The dust mass can be determined
from the emission spectrum with some assumptions on the dust
emissivity and composition.  Adopting a gas-to-dust ratio the dust
mass can be used to estimate the molecular gas mass.  This is however
not a trivial business because the gas-to-dust ratio is derived
locally in the Milky Way and may vary for different conditions.  The
CO abundance strongly depends on the amount of dust, because the lower
the amount of dust, the lower the shielding of molecules, hence they
are easier to dissociate, which in turn will render CO as an
unreliable tracer of the molecular gas.

To study the relation of cold dust and molecular gas in low
metallicity environments we have
therefore decided to observe \object{NGC\,4449} with the Submillimetre
Common-User Bolometer Array (SCUBA) at the James Clerk Maxwell
Telescope (JCMT) at 850 and 450~$\mu$m.
Maps of the cold dust in the millimetre and submillimetre
regime for dwarf galaxies are rather rare.
Together with the investigation of \object{NCG\,1569} (Lisenfeld et
al. \cite{2002A&A...382..860L}) the study of \object{NGC\,4449}
presented here is the second  of its kind.

The galaxy \object{NGC~4449} has seen numerous studies over
essentially the whole electro-magnetic spectrum (see e.g. Hunter et
al. \cite{1999AJ....118.2184H}, Hill et al.
\cite{1998ApJ...507..179H}, Klein et al. \cite{1996A&A...313..396K}).
It is a nearby Magellanic irregular dwarf galaxy of high surface
brightness at a distance of about 3.7~Mpc (Bajaja et
al. \cite{1994A&A...285..385B}). The recent investigation of
Karachentsev et al. (\cite{2003A&A398..467}) yielded a distance of
4.12$\pm$0.50~Mpc, which agrees with the older value within the
errors.  Located in the constellation Canes Venatici it is a member of
the CVnI group (Sandage \cite{1975gaun.book.....S}), which contains
mainly spirals and irregular galaxies.  The observed morphology and
kinematics of the HI gas suggests a tidal interaction with the nearby
dwarf galaxy DDO~125 (Theis \& Kohle \cite{theis}) and may explain the
high star-formation rate and the huge HI streamers surrounding
\object{NGC\,4449}. The metallicity was found to be
12+log(O/H)~=~8.3 (Lequeux et al., \cite{leque}).

Hunter et al. (\cite{1986ApJ...303..171H}) stated that the spectral
distribution of the far-infrared emission from \object{NGC\,4449} as
observed with IRAS is not different from normal spirals.  The ratio of the
IRAS flux densities at 100 and 60$\mu$m
\mbox{$S_{\rm 100\mu m}/S_{\rm 60\mu m}\approx2$} is at the lower
end of the range covered by spirals and indicates a high star
formation rate.  However, the dust mass calculated from this
measurement only relies on the warm dust components, thus representing
a lower limit.  Our new observations allow to determine the amount
of cold dust in \object{NGC\,4449} and to study the gas-to-dust ratio in
more detail.  In Sect.~2 we describe the observations and data
reduction procedure. Sect.~3 presents the results, while in Sect.~4 a
comparison with the gas in
\object{NGC\,4449} is made. A summary is given in Sect.~5.

 \section{Observations and data reduction}

\object{NGC\,4449} was imaged simultaneously at 850 and 450~$\mu$m
using the SCUBA camera at the JCMT in January 18, 1999. The
observations were made in the jiggle mode with a chopper throw of
120\arcsec. Two fields of view of about 2\farcm3 with sufficient
overlap were observed. The extent of the galaxy at optical wavelengths
is about 2\arcmin\ by 3\arcmin.  Since we observed about 9 hours with
the source rotating on the sky, there may have been some emission from
the galaxy itself in the reference beam occasionally. However, by
combining all coverages this error should be small compared to the
calibration uncertainties. Thus only a diffuse extended dust
component possibly associated with the large HI halo
would have been missed by our observations.

The total net observing time was 8960~s for
field A and 6528~s for field B.  The two fields were placed around a
central position at \mbox{$\alpha_{\rm 1950}$=12$^h$25$^m$46$^s$,
$\delta_{1950}$=44\degr22\arcmin45\arcsec} such as to cover the main
(CO emitting) body of the galaxy.  The whole observing run (9~hours in
total) had good and stable weather with an optical depth of
\mbox{$\tau_{\rm 850}\approx0.23$} and a seeing of \mbox{$\sim0\farcs4$}.
Flux calibration was made using Mars, which was observed
in the middle and at the end of the observing run.

\begin{figure*}
\resizebox{15.5cm}{!}{\includegraphics{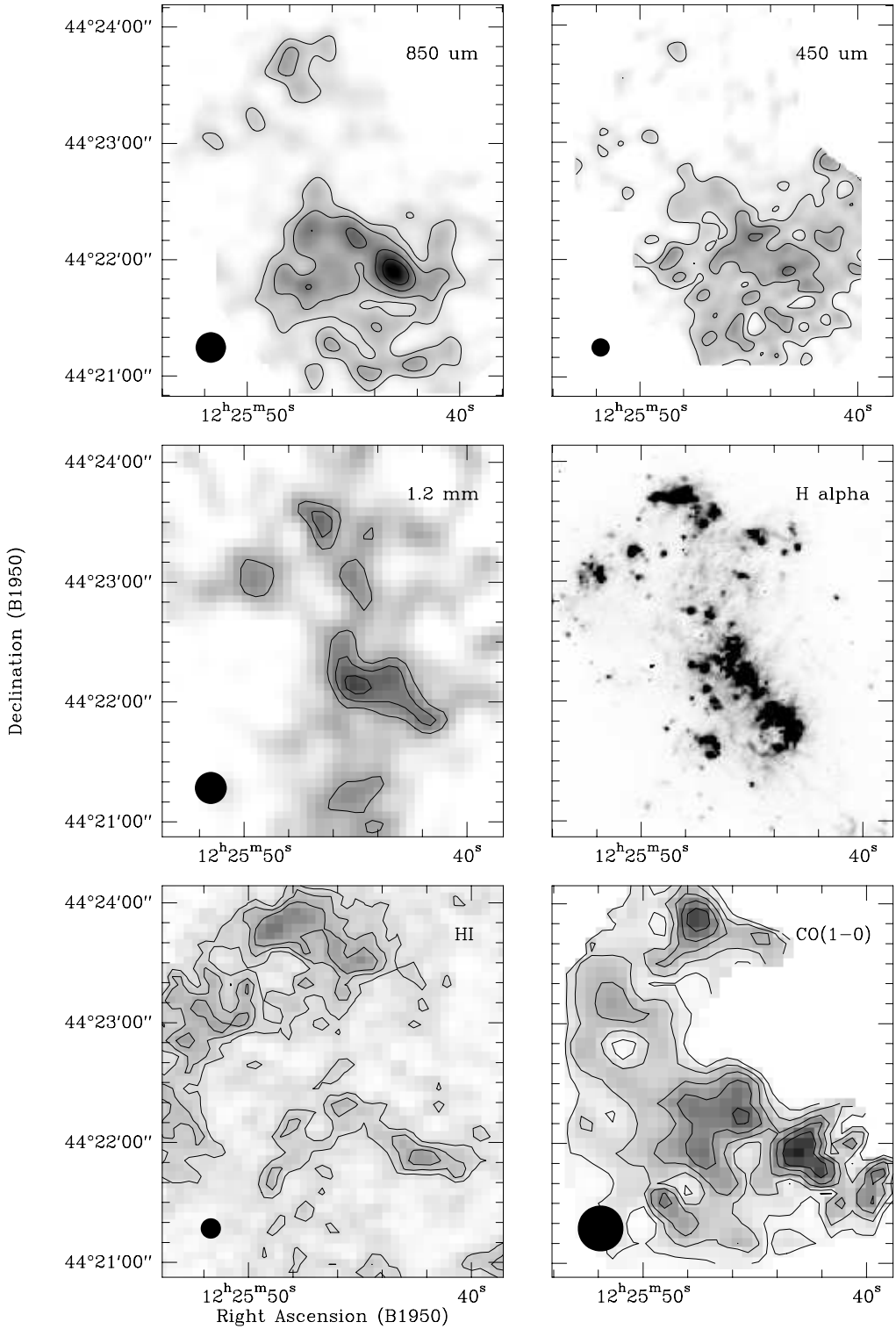}}
\caption{Intensity maps of \object{NGC\,4449} at 850
(upper left) and 450~$\mu$m (upper right), presented in this work,
with contour levels of 2,3,4 and 5~$\sigma$, where $\sigma$ is the
noise level of the map
($\sigma(450{\rm\mu m})=18$mJy/beam and
$\sigma(850\mu{\rm m})=6$mJy/beam); at 1.2~mm (centre left, from Kohle
\cite{kohle99}), with contour levels of 2, 2.5 and 3~$\sigma$
($\sigma(1200\mu{\rm m})=2$mJy/beam); an
H$\alpha$ image (centre right, from Bomans et
al. \cite{1997AJ....113.1678B}), the integrated HI (bottom left, from
Hunter \cite{1999AJ....118.2184H}), with contour levels of 0.06, 0.08 and
0.1~Jy/beam and the integrated CO(1-0) (bottom right, from Kohle et al.
\cite{kohle98}, \cite{kohle99}), with contour levels from 0.5 to
2.5~K~km~s$^{-1}$ in steps of 0.5~K~km~s$^{-1}$. All maps show the same
area as that mapped with
SCUBA.}
\label{all}
\end{figure*}

The data have been reduced in the standard manner using the STARLINK
reduction package SURF (Jenness \& Lightfoot
\cite{1999StarlinkUserNote216}).  Flatfielding was performed to
correct for differences in the sensitivity of the bolometers.  The
atmospheric opacity was determined from skydip observations.  Noisy
bolometers and spikes were identified and blanked out.  Typical
pointing corrections were 1\arcsec\ to 2\arcsec, hence small compared to the
beam size.
From the planet profile, the beam size (HPBW) was estimated to
15\farcs3 and 8\farcs5 at 850 and 450~$\mu$m, respectively.
Fig.~\ref{all} shows the resulting maps of \object{NGC\,4449} at 850
and 450~$\mu$m in the top panels.  The noise levels of the maps are
\mbox{$\sigma$~=~6~mJy/beam} at 850~$\mu$m and
\mbox{$\sigma$~=~18~mJy/beam} at 450~$\mu$m.

Observations at 1200~$\mu$m were previously carried out with the
IRAM~30~m telescope on Pico Veleta (Spain) in December 1998, using the
19-channel bolometer array of the Max-Planck-Institut f\"ur
Radioastronomie (Kohle \cite{kohle99}). We obtained On-The-Fly
(OTF) maps with sizes between 300\arcsec$\times$200\arcsec\ and
400\arcsec$\times$300\arcsec\ in azimuth and elevation, with the
telescope moving in azimuth at a speed of \mbox{6\arcsec\ /s}. Successive
horizontal scans were spaced by 4\arcsec\ in elevation. The wobbler
throw was 45\arcsec\ and wobbled at 0.5~Hz in azimuth.
During the run the weather conditions were good
and the atmosphere stable, with sky opacities between 0.1 and 0.3 at
230~GHz.  Two channels had to be discarded. The total integration time
was about 8 hours.  Calibration was made using Mars, which was
observed twice during the run.  The derived beam-size is 10\farcs8 and
the resulting map is shown on the left of the centre row in
Fig.~\ref{all}, smoothed to match the 15\farcs3 of the 850~$\mu$m map
and to improve the signal-to-noise. The noise level of this map is
\mbox{$\sigma$~=~2~mJy/beam}.

We have to mention that restoring maps from chopped on-the-fly
measurements may cause photometric imperfections, as the information on
the large-scale structure (larger than the wobbler throw) is lost. The
same holds true of course for the IRAM data: we may have missed a diffuse,
extended dust component and hence a noticeable fraction of the integrated
flux.
Such a component was suggested by Thronson et al. (\cite{1987ApJ...317..180T})
which detected emission at 150~$\mu$m with the Kuiper Airborne Observatory (KAO)
on scales of up to 4 arcmin at a low signal-to-noise ratio.
However, this will not change our general conclusions. In particular,
if we would be able to correct for a missing large-scale component, we
would flatten the slope of the submm spectrum, and increase the dust mass
(see Sect.~3).

 \section{Results}

Fig.~1 exhibits the dust continuum emission of \object{NGC\,4449}
resulting from our observations, along with images in the HI,
H$\alpha$ and CO line.  The galaxy is clearly detected at all mm/submm
wavelengths.  The morphology of the galaxy is rather similar for all
components, except for the neutral atomic hydrogen.  The strongest
dust emission comes from a bar-like structure prominent in the optical
regime, although the brightest regions do not coincide.  The optically
bright northern region is conspicuous in all maps and manifests a
large star-forming complex which is most pronounced in the H$\alpha$
map.  A more detailed comparison is hardly possible, due to the low
signal-to-noise and therefore the uncertainties in the exact
structures.

By integrating over the whole SCUBA maps we have determined total flux
densities to be $S_{\rm 850\mu m}=1.8\pm0.2$~Jy and
$S_{\rm 450\mu m}=16.5\pm4.1$~Jy. No color correction has been made.
Because the CO(3-2)-line emission
falls into the bolometer bandpass at 850~$\mu$m, we corrected the
total continuum flux for the line contribution.  We estimated the
contamination using the CO(1-0) map and assuming a line ratio of
\mbox{$I_{\rm CO(3-2)}/I_{\rm CO(1-0)}\approx0.7$} (Fritz 2002, private
communication).
We thus obtained a contribution of the CO line to the total emission
of $\sim$~12\%, resulting in a pure continuum flux of
1.5~$\pm$~0.3~Jy.  The contamination at 450~$\mu$m is negligible
compared to the strong thermal dust emission, hence no correction was
made.  The integrated flux at 1.2~mm was taken from Kohle
(\cite{kohle99}) and is also corrected for contributions of the
CO(2-1) line.
The quoted errors are formal errors based on the map
noise and on the uncertainty of integrating the flux, including
calibration errors.

\begin{figure}
\resizebox{\hsize}{!}{\rotatebox{270}{\includegraphics{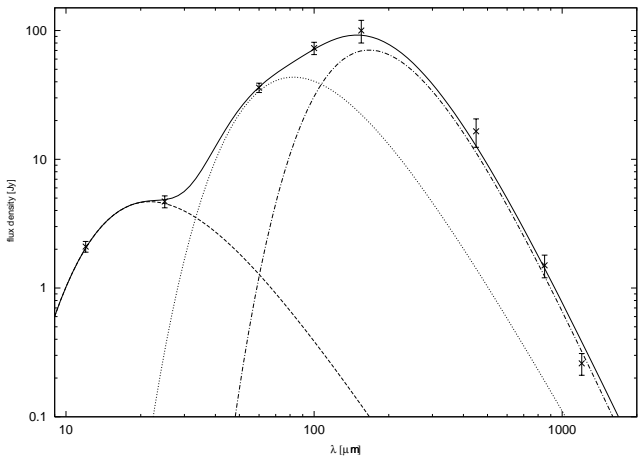}}}
\caption{The mid-infrared to millimeter spectrum of \object{NGC\,4449}, fitted
with three-temperature dust components. Crosses mark flux densities from
Table~1. The three dust components have temperatures of 168~K (dashed line),
39~K (dotted line) and 16~K (dashed-dotted line).}
\label{threetemp}
\end{figure}

In the literature there are 5 published FIR flux densities for
\object{NGC\,4449}. These are the four IRAS data points between
12~$\mu$m an 100~$\mu$m (Hunter et al.
\cite{1986ApJ...303..171H}), and a 150~$\mu$m value from KAO observations
(Thronson et al. \cite{1987ApJ...317..180T}).  Together with
the data points presented in this paper this leads to a fair
sampling of the spectral energy distribution of \object{NGC\,4449}
from mm through FIR wavelengths.  In Table~1 we have compiled all
values, Fig.~\ref{threetemp} showing the resulting spectrum.

\begin{table}
\caption{IR to submm integrated flux densities of \object{NGC\,4449} from
12~$\mu$m to 1200~$\mu$m. The values at 850 and 1200~$\mu$m have been corrected
for contamination by CO lines.}
\begin{center}
\begin{tabular}{rcc}
\hline
$\lambda$ & flux density & Ref. \\
($\mu$m)  &(Jy) & \\
\hline
12       & 2.1   $\pm$ 0.2   &  Hunter et al.   (\cite{1986ApJ...303..171H})\\
25       & 4.7   $\pm$ 0.5   &  Hunter et al.   (\cite{1986ApJ...303..171H})\\
60       & 36    $\pm$ 3     &  Hunter et al.   (\cite{1986ApJ...303..171H})\\
100      & 73    $\pm$ 8     &  Hunter et al.   (\cite{1986ApJ...303..171H})\\
150     & 100   $\pm$ 20    &  Thronson et al. (\cite{1987ApJ...317..180T})\\
450      & 16.5  $\pm$ 4.1   &  this work\\
850      & 1.5   $\pm$ 0.3 *  &  this work\\
1200     & 0.26  $\pm$ 0.04 * &  Kohle           (\cite{kohle99})\\
\hline
\end{tabular}
\end{center}
\begin{center}
* corrected for CO line contamination \\
\end{center}
\end{table}

If we assume the dust grains to be in equilibrium with the radiation
field we can describe their emission by a modified black-body law.
This will definitively be the case for large grains, which contribute
most to the sub-millimetric continuum.  They also account for more
than 90\% of the dust mass (see e.g. D\'esert et al. \cite{1990A&A237..215}).
The emission from 20 to 80~$\mu$m can originate from very small grains
at a significant amount. In this case, with our simple approach we may
overestimate the mass of that component by a few percent. However, as we
will see below, the mass of this hot component is of almost no importance.

In case of a low optical depth the intensity is given by
\mbox{$F_\lambda = B_{\lambda}(T) \cdot k(\lambda)$}, where
$B_{\lambda}(T)$ is the Planck function, $T$ the dust temperature and
$k(\lambda)$ is the absorption coefficient.  The wavelength dependence
of the latter, which is usually assumed to follow a power-law
\mbox{$k(\lambda)\propto\lambda^{-\beta}$} (see e.g. Alton et al.
\cite{2000A&A356..795}), is not precisely known.
Models of the interstellar dust yield a spectral index
$\beta=1\ldots2$ for mixtures of small and big grains (e.g. Draine \&
Lee \cite{draine}). Gordon (\cite{gordon}) has noted a temperature
dependence of $\beta$, for the range $1\ldots2.5$, and Dunne \& Eales
(\cite{2001MNRAS.327..697D})
have derived $\beta\simeq2$ on a statistical basis for a sample
of bright IRAS galaxies.

In order to further analyze the dust components of \object{NGC\,4449}
we have modeled the spectrum with three dust components at different
temperatures.
This yields a reasonable fit to the emission spectrum over the
entire observed wavelength range.  As we have only eight observed
spectral data points, we have to limit the model to three components.
In reality, the dust is probably at a range of temperatures,
reflecting the range of colours and intensities of the heating radiation fields.
We decided to use $\beta=1$ for the hot and $\beta=2$ for the warm component,
leaving 7 free parameters for the fit. It was not possible to fit the steep
slope of the spectrum satisfactorily with $\beta=2$ for the cold
component, thus we decided to leave it free up to 2.5, resulting in a
value of $\beta=2.3$.  Fig.~2 shows the result with hot
\mbox{($T=168\pm10$~K, $\beta=1$)}, warm \mbox{($T=39\pm3$~K,
$\beta=2$)} and cold \mbox{($T=16\pm2$~K, $\beta=2.3$)} dust
components.

The derived parameters are in accord with other investigations
(e.g. Dumke et al. \cite{dumke}), and agree well with measurements of
the diffuse ISM in the Galaxy \mbox{($T\approx17$~K, $\beta=2$)} by
COBE/FIRAS (Reach et al. \cite{1995apj451..188}, Sodroski et
al. \cite{1997apj480..173}).  In order to properly fit the relatively
strong (sub)millimeter emission, a cold component is required.
Moreover, because of the low dust emissivity at this temperature, a
very large fraction of the total dust has to be cold in order to
explain the emission at 850~$\mu$m and 1200~$\mu$m.  An attempt to fit
a two-temperature model yields significantly worse results.

From the three-component fit to the spectrum of \object{NGC\,4449} we
can estimate the dust mass.  This can be achieved with a simple
equation adapted from Hildebrand (\cite{hildebrand}), where

\begin{equation}
M_{\rm d}= \frac{D^2 S_{\rm 850 \mu m}}{B_{\rm 850 \mu m}(T) k_{\rm 850 \mu m}}.
\label{mdust}
\end{equation}

Here, $B_{\rm 850 \mu m}(T)$ is the Planck function at 850~$\mu$m, $D$ is
the distance and $S_{\rm 850 \mu m}$ the observed flux density.
To derive a consistent
value of the dust absorption coefficient $k_{\rm 850 \mu m}$ we used the
above mentioned power-law \mbox{$k(\lambda)\propto\lambda^{-\beta}$}
and the canonical value of
\mbox{$k_{\rm 250 \mu m}=10$ ~cm$^2$~g$^{-1}$} from Hildebrand
(\cite{hildebrand}), extrapolating to the relevant
wavelengths and using the associated value of $\beta$.  We thus
obtained the following values of the dust absorption coefficient for
the three different components:
\mbox{$k_{\rm 850 \mu m}=0.6$ ~cm$^2$~g$^{-1}$} for the cold (16~K)
component at 850~$\mu$m, \mbox{$k_{\rm 100 \mu m}=63$~cm$^2$~g$^{-1}$} for
the warm (39~K) and \mbox{$k_{\rm 100 \mu m}=25$~cm$^2$~g$^{-1}$} for the
hot component at 100~$\mu$m.  These values are consistent given our
assumptions, and especially for the spectral indices used here.
However, in the literature one can find a wide range of values for the
emissivity and the dependences are only poorly known.  As the most
extreme cases we quote two values, namely
\mbox{$k_{\rm 850 \mu m}=0.26$ ~cm$^2$~g$^{-1}$}
(Alton et al. \cite{2002A&A388..446}) and
\mbox{$k_{\rm 850 \mu m}=2.4$ ~cm$^2$~g$^{-1}$}
(Alton et al. \cite{2000A&A356..795}).

The three components shown in Fig.~\ref{threetemp} result in a very
small mass of hot dust, \mbox{$M_{\rm d,h}\simeq18$~{M$_\odot$}}, a larger
mass of warm dust, \mbox{$M_{\rm d,w}=0.3\times10^5$~{M$_\odot$}} and a large
mass of the cold dust, \mbox{$M_{\rm d,c}=3.8\times10^6$~{M$_\odot$}}. The
total dust mass, \mbox{$M_{\rm d}=3.8\times10^6$~{M$_\odot$}}, is thus
completely dominated by the cold dust component.
The uncertainty of this mass is naturally large
(within a factor of 3), and is mainly governed by the peak and the
slope of the spectrum between \mbox{100~-~1200~$\mu$m}, and by the
assumed dust absorption coefficient, which is not well known.  The
error in calculating the dust mass caused by using modified Plankians is
negligible in view of the large uncertainties in the general grain
properties.

Thronson et al. (\cite{1987ApJ...317..180T}) calculated a somewhat
lower mass of \mbox{$M_{\rm d}\approx1.5\times10^6$~{M$_\odot$}}, using
their 150~$\mu$m value and assuming a spectral index of $\beta=1$; the
deviation to our value is most likely caused by the then
unknown slope of the spectrum in the millimetre regime.

Lisenfeld et al. (\cite{2002A&A...382..860L}) have undertaken a
similar study of \object{NGC\,1569}, a galaxy which is presently in
the aftermath of a massive burst of star formation. The dust
spectrum of that galaxy shows a completely different slope in the
range between \mbox{150~-~1200~$\mu$m}. The IRAS-flux densities of
\object{NGC\,1569} indicate a different location of the maximum
between 100 and 60~$\mu$m, compared to \object{NGC\,4449}, indicating
the presence of much more hot dust in the former. A very cold
component of around 7~K was needed to properly fit the spectrum,
which resulted in a tremendous mass of this very cold dust.
The authors stated that such a large amount of cold dust is very unlikely
in a dwarf galaxy dominated by intense radiation fields
(2 orders of magnitude higher than in the solar neighbourhood).
Hence, Lisenfeld et al. favour the model of D\'esert et al. (\cite{1990A&A237..215}),
where the very small grains have optical properties consistent with beta=1. Since
\object{NGC\,1569} has a shallow slope, the resulting fit requires a large contribution of
grains emitting like beta=1, thus a large contribution of VSG with the D\'esert's
et al. properties.
The dust mass derived from this model is much less than that from a fit with
modified Plankians and results in a overall gas-to-dust ratio of
\mbox{$M_{\rm g}/M_{\rm d}\approx~1500~\cdot\cdot\cdot~2900$}. We
have to point out, however, that the physical reason for the steeper
slope of the submm dust continuum spectrum of NGC\,4449 ($\beta=2.3$
vs. $\beta=1.0$ in case of NGC\,1569) is not at all obvious.

 \section{Comparison with the gas}

The derived dust distribution can be compared with other tracers of
the ISM, viz. CO and HI. We also have a CO(1-0) map of
\object{NGC\,4449} at our disposal, obtained with the IRAM 30-m
telescope on Pico Veleta (Kohle \cite{kohle98}, \cite{kohle99}).
Fig.~\ref{all} shows
the integrated map, the angular resolution is 22\arcsec .  The
correlation of CO and dust is clearly seen.  However, the CO coverage
of the galaxy is not uniform, especially in the outer parts so that
not all of the structures seen there are reliable.

We have also compared the H$\alpha$ map published by Bomans et al.
(\cite{1997AJ....113.1678B}) and shown in Fig.~\ref{all} to the maps
of the dust and the molecular gas. It is evident that the strongest
dust emission follows the distribution of intense star formation (as
expected). However, this is not a simple one-to-one relation. The same
is true for the CO emission and may indicate dissociation processes as
observed in massive star forming regions. There is also significant
dust emission away from the star-forming regions, but this is
accompanied by CO radiation and may show a more diffuse component
(smaller clouds) of the molecular gas.

A comparison with the HI (Hunter et al. \cite{1999AJ....118.2184H}) shows
only little correlation. The inner HI of NGC\,4449 is organized in a
ring-like feature around the optical body. In the northeast one can
see part of this ring, whereas in the south there is nearly no HI
within our mapsize.  In the center of \object{NGC\,4449} only little
HI is found, which is expected. The gas should be ionized there, in
molecular form or frozen out onto the dust grains.

With the dust mass of \mbox{$M_{\rm d}=3.8\times10^6$~{M$_\odot$}} derived above we
can estimate the total gas mass. Using the normal gas-to-dust ratio of
\mbox{$M_{\rm g}/M_{\rm d}\approx$~130} (Devereux \& Young \cite{dever})
this yields a total gas mass of \mbox{$M_{\rm HI+\rm H_2}=4.9\times10^8$~{M$_\odot$}}
for the inner (optical) part of the galaxy.
Computing the HI mass within our mapsize yields \mbox{$M_{\rm HI}=1.5\times10^8$~
{M$_\odot$}}.
Subtraction from the total gas mass
estimated above results in a mass of molecular hydrogen of
\mbox{$M_{\rm H_2}=3.4\times10^8$~{M$_\odot$}}. From the integrated CO spectrum we
derive a CO luminosity of
\mbox{$L_{\rm CO}=7.9\times10^6$~K~km~s$^{-1}$~pc$^2$},
which leads to a conversion factor from CO intensity to H$_2$ column density
\mbox{{$X_{\rm CO}$}$=25\times10^{20}$~cm$^{-2}~($K~km~s$^{-1})^{-1}$},
i.e. about 11 times
the standard Galactic value quoted in Sect.~1. Such an enhanced value
is explicable in view of the lower metallicity, of \object{NGC\,4449},
viz. 12+log(O/H)~=~8.3 (Lequeux et al., \cite{leque}).

Kohle (\cite{kohle99}) derived an even higher {$X_{\rm CO}$}, assuming virialized cloud
complexes in the CO map. He obtained
\mbox{$X_{\rm CO}=(37\pm11)\times10^{20}$~cm$^{-2}~($K~km~s$^{-1})^{-1}$},
 which is about 16 times the Galactic value and may represent an upper
 limit. This would lead to a much higher molecular gas mass and hence
 to an even higher gas-to-dust ratio. Both scenarios are in agreement
 with what has been found for other low-mass dwarf irregular galaxies
 (see e.g. Barone et al., \cite{barone} and references therein).
 Thronson et al. (\cite{1987ApJ...317..180T}) derived
\mbox{$X_{\rm CO}/X_{\rm MW}~\approx~10$} from their KAO data.
Based on a much poorer sampling of the dust spectrum using IRAS CPC
data Israel (\cite{israel}) calculated a value of
\mbox{$X_{\rm CO}=8\times10^{20}$~cm$^{-2}$~(K~km~s$^{-1})^{-1}$}, which is
significantly lower than the other values discussed here.

 \section{Conclusions}

We have presented new maps of the dwarf galaxy \object{NGC\,4449} at 450
and 850~$\mu$m, taken with SCUBA at the JCMT. The integrated flux densities
were compared to those obtained with IRAS and KAO.
The derived flux densities can be readily explained by the presence of a
significant amount of cold dust. We were able to fit the observed SED well
with modified Plankians of three dust components, requiring most of the dust
to be at a temperature of 16~K, with a spectral index $\beta=2.3$, which is
a bit higher than the frequently used value of 2.

Comparison with CO observations shows a good coincidence of the cold dust and
the molecular gas. The H$\alpha$ image shows that star-forming regions are
related to the dust emission peaks. The main emission comes from the bar-like
structure, but there is also weak dust emission away from the main body of
\object{NGC\,4449}, and there may be some more following the HI ring-structure.

Assuming a standard local gas-to-dust ratio, we derive a conversion factor
\mbox{$X_{\rm CO}\approx25\times10^{20}$~cm$^{-2}~($K~km~s$^{-1})^{-1}$}, which is 11
times the Galactic value. Given the low metallicity of
\object{NGC\,4449}~(1/3~solar), this is probably still a lower limit,
because a standard gas-to-dust ratio may not be applicable. The
standard Galactic conversion factor thus strongly underestimates the
molecular gas mass in such galaxies, because our tracer -- the CO
molecule -- is less abundant, due to the lower metallicity and/or
dissociation. In view of the low metallicity of \object{NGC\,4449}
this seems in perfect agreement with studies of other dwarf
irregulars.

\begin{acknowledgements}
We are grateful to Gerald Moriarty-Schieven for performing the observations and to
D. A. Hunter for making the HI data available to us. We are also indebted to C.
Kramer and the anonymous referee for many helpful suggestions.

\end{acknowledgements}

\end{document}